\documentstyle[emulateapj,onecolfloat]{article}
\input epsf

\renewcommand{\bv}{{\bf v }}
\newcommand{\be}{{\bf e }}
\newcommand{\tdyn}{t_{\rm dyn}}

\begin{document}

\twocolumn[
\title{A Quantitative Study of Interacting Dark Matter in Halos}
\author{C.S.~Kochanek and Martin White}
\affil{Harvard-Smithsonian Center for Astrophysics, Cambridge, MA 02138}
\affil{ckochanek, mwhite@cfa.harvard.edu}

\begin{abstract}
\noindent
\rightskip=0pt
We study the evolution of Hernquist profile ``galaxies'' in the presence of
self-interacting dark matter (SIDM), where the properties of the dark matter can
be parameterized by one number, $\hat{\sigma}_{DM}=\sigma_{DM} M_T/a^2$ for
a halo of mass $M_T$ and break radius $a$.  While the halos form constant
density cores of size $\sim a/2$ on the core radius relaxation time scale
($t_{rc} \simeq 1.7\tdyn/\hat{\sigma}_{DM}$) core collapse begins shortly
thereafter and a steeper $1/r^2$ central density cusp starts forming faster
than predicted by 2-body relaxation. The formation of the steeper central cusp is
accelerated if the cooling baryons adiabatically compress the dark matter.  
The natural consequence of SIDM is to exacerbate rather than to mitigate
astrophysical problems created by dark matter density cusps.
\end{abstract}
]

\section{Introduction}

A model based on the growth of small fluctuations through gravitational
instability in a universe with cold dark matter (CDM) provides an excellent
fit to a wide range of observations on large scales ($\gg 1$Mpc).
However the nature and properties of the CDM, apart from its being cold and
dark, remain mysterious.  It is one of the major goals of cosmology to
further constrain the properties of the dark matter and determine its nature.
Recently much attention has been refocused on this question because of a
suggestion by Spergel \& Steinhardt~(\cite{SpeSte}) that possible discrepancies
with observations on kpc scales could be probes of dark matter properties,
particularly its self-interaction rate.
In its simplest incarnation, the self-interacting dark matter (SIDM) model
provides a 1-parameter family of models with standard CDM as a limiting case,
and it is thus interesting to examine this model in some detail.
 
In this paper we calculate the quantitative effects of SIDM on the evolution of 
isolated halos using an N-body code with particle scattering.
We review the astrophysical arguments motivating SIDM in \S\ref{sec:motivate}
and describe our implementation of the scattering algorithm in
\S\ref{sec:method} (our method is very similar to that of Burkert~(\cite{Bur}),
but uses a more accurate approximation to the scattering rate integral).
The impact of self-interaction on various astrophysical processes is discussed
in \S\ref{sec:core} and we finish with some conclusions in
\S\ref{sec:conclusion}.

\section{The case for SIDM} \label{sec:motivate}

In this section we briefly recap the arguments which led 
Spergel \& Steinhardt~(\cite{SpeSte}) to revive the SIDM model
(Carlson et al.~\cite{CMH}; de Laix, Scherrer \& Schaeffer~\cite{deLaix})
since it is precisely these astrophysical observations that can be used to
constrain the self-interaction of dark matter.

There is a growing consensus that CDM halos form with central density cusps,
where $\rho\sim r^{-\gamma}$ as $r\rightarrow 0$ with $\gamma\simeq 1$
(Navarro et al.~\cite{NFW}; Moore et al.~\cite{Moo98};
Huss et al.~\cite{Huss}; Klypin et al.~\cite{Kly}; Moore et al.~\cite{Moo99};
but see Kravstov et al.~\cite{Kravstov}).
Analytic arguments (Syer \& White~\cite{SyeWhi}; Kull~\cite{Kull}) suggest
that such a core structure follows naturally from merging in collisionless
systems.
These results are (in some senses) confirmed experimentally by the similar
stellar density cusps found in early-type galaxies
(e.g.~Faber et al.~\cite{Faber}), which also largely formed through the
merging of collisionless matter (stars).  However, such density cusps are
inconsistent with the finite core radius required to reproduce the rotation
curves of dwarf galaxies (Moore~\cite{Moo94}; Flores \& Primack~\cite{FloPri}).
The problem appears to be limited to very low rotation speed galaxies
($v_c<10^2$~km~s$^{-1}$) as van den Bosch et al.~(\cite{vdBRobDalBlo}) have 
now shown that the slightly more massive LSBs (low surface brightness galaxies) 
actually {\it require\/} cusped dark matter profiles.  CDM simulations also
produce halos which correctly reproduce the numbers of galaxies in clusters,
but at first sight grossly overpredict the number of satellites orbiting the
Galaxy (Moore et al.~\cite{MGGLQST}).

One solution to the discrepancies is to blame the differences on the
effects of the baryons.  The dwarf galaxies are the systems where the baryons
can most easily modify the core structure of the dark matter (e.g.~Hernquist
\& Weinberg~\cite{HerW}; Navarro 
Eke \& Frenk~\cite{NEF}; Gelato \& Sommer-Larsen~\cite{GSL};
Binney, Gerhard \& Silk~\cite{BinGer}) although it is difficult to
quantitatively estimate the effects of feedback
(e.g.~MacLow \& Ferrara~\cite{Maclow}).
Similarly, massive spiral galaxies are probably the best environment for the
effects of the baryons to modify the satellite predictions of pure dark matter
simulations (e.g.~Bullock, Kravstov \& Weinberg~\cite{BKW}).
It is, however, unclear whether the baryons can provide a complete solution to
the two problems.  A second solution is to use hot rather than cold initial
conditions for the collapsing dark matter (e.g. Hozumi, Burkert \& Fujiwara~\cite{Hozumi}).

Self-interacting dark matter may be a third solution, because it can also
heat the cores of dark matter halos and evaporate orbiting satellites given
an appropriately tuned interaction cross section.  This idea has attracted 
considerable attention
(Ostriker~\cite{Ost}; Hannestad~\cite{Hann}; Hogan \& Dalcanton~\cite{HogDal};
Burkert~\cite{Bur}; Firmani et al.~\cite{FOACH}; Mo \& Mao~\cite{MoMao};
Hannestad \& Scherrer~\cite{HanSch}, Bento et al.~\cite{Bento}),
but the detailed quantitative consequences of the theory have yet to be
derived.  We have used N-body methods, discussed in the next section, to
explore the formation and evolution of ``galaxies'' in a simple model.

\section{Numerical Methods} \label{sec:method}

Throughout we shall study $10^5$ particle realizations of a Hernquist
profile (Hernquist~\cite{Her})
\begin{equation}
  \rho_H(r) = {M_T\over 2\pi} {a\over r} {1\over(r+a)^3}
\end{equation}
which has two parameters: the total mass $M_T$ and break radius $a$.
The orbital time at the break radius $a$ is $\tdyn=4\pi (a^3/GM_T)^{1/2}$,
which we shall sometimes refer to as the ``dynamical time''.
With $10^5$ particles we can probe the halo structure down to $\sim0.1a$,
which is sufficient for our purposes.
We evolve the system using a {\sl Tree\/} code described in
Appendix~\ref{sec:tree}.

In addition to the usual gravitational force, we implement the 
self-interaction of the dark matter using a Monte-Carlo technique.
First let us define some dimensionless units.
For a galaxy of mass $M_0$ and radius $R_0$ (where $M_0=M_T$ and
$R_0=a$ for the Hernquist model) define a density scale
$\rho_0=M_0/R_0^3$, a velocity scale $v_0=(GM_0/R_0)^{1/2}$ and a
time scale $t_0=R_0/v_0$.  
The simplest self-interaction has the scattering cross section per unit mass
of the dark matter, $\sigma_{DM}$, independent of energy.  In this case we
obtain the 1-parameter family of models which we shall study here.
The scattering rate for a particle with velocity $\bv_0$ is
\begin{equation}
   \Gamma = { d n \over d t} = \int d^3 \bv_1 f(\bv_1) \rho  \sigma_{DM}
    |\bv_0-\bv_1|
\end{equation}
where $f(\bv_1)$ is the local distribution function of velocities.
In dimensionless units, 
\begin{equation}
  \hat{\Gamma} = { d n \over d \hat{t} } = \hat{\sigma}_{DM} 
  \int d^3 \hat{\bv} f(\hat{\bv}_1) \hat{\rho} |\hat{\bv}_0-\hat{\bv}_1|
  \label{eqn:scat}
\end{equation}     
with dimensionless cross section $\hat{\sigma}_{DM}=M_0\sigma_{DM}/R_0^2$.
For a Hernquist model
the fraction of the particles interacting per crossing time $t_0$ is
$0.05\hat{\sigma}_{DM}$, the mean free path is
$a/\hat{\rho}\hat{\sigma}_{DM}$, and these interactions have a
typical radius $\sim a/2$.
We shall quantify the effect of self-interaction as a function of
$\hat{\sigma}_{DM}$, and one can then scale to cases of astrophysical
interest by suitable choice of the basic dimensional parameters
$M_0$ and $R_0$.

We have simulated dimensionless cross sections of 
$\hat{\sigma}_{DM}=M_T\sigma_{DM}/a^2=0$, $0.3$, $1.0$, $3.0$ and
$10.0$.
For $\hat{\sigma}_{DM}=1$ we also ran a case with $8 \times 10^5$ particles,
doubling the spatial resolution, and a case using the Burkert~(\cite{Bur}) 
scattering algorithm for comparison.
The evolution of Navarro, Frenk \& White~(\cite{NFW}) models should be
similar if we scale the two models to have the same central density
distribution.  For this normalization, the evolution of an NFW model will
match that of our Hernquist models if
$\hat{\sigma}_{DM} = 2\pi\delta_c \rho_{crit} r_c \sigma_{DM}$. 

To implement the scattering in an N-body code we need to discretize the
calculation of the integral to compute the scattering rate for each particle.
Call the particle whose rate we are calculating $0$ and particle $j$ the
particle from which it is scattering.
We approximate the local density from the total mass and radius enclosed by
the $N=32$ nearest neighbors which we order in increasing distance from
particle $0$.  Writing the masses and velocities of the neighbors as
$\hat{m}_j$ and $\hat{\bv}_j$ with $j=1,\cdots,N$ we estimate the scattering
rate for particle $0$ by 
\begin{equation}
  \hat{\Gamma} = \hat{\sigma}_{DM} \hat{\rho}_N { \sum_{j=1}^N \hat{m}_j 
                 |\bv_j-\bv_0|  \over \sum_{j=1}^N \hat{m}_j }
\end{equation}
where
\begin{equation}
  \hat{\rho}_N =  { 3 \sum_{j=1}^N \hat{m}_j \over  4\pi r_N^3}
\end{equation}
is the estimate of the local density.
Note that this differs from the prescription of Burkert~(\cite{Bur}) who
simply approximated $|\Delta \bv|=|\bv_0|$, thereby underestimating the
heating rate for low velocity particles.

For time step $\Delta\hat{t}$ the probability of an interaction during the
time step is $P_{\rm int}=\hat{\Gamma}\Delta\hat{t}$.  We adjust the time
steps to keep the interaction probability relatively low ($P_{\rm int}<0.5$),
and the actual number of interactions in each time step is determined from 
the Poisson distribution with an expectation value of $P_{\rm int}$.  
For each scattering event we interact with one of the $N$ nearest particles,
where the probability of interacting with particle $j$ is 
\begin{equation}
  p_j = { m_j |\bv_j-\bv_0| \over  \sum_{i=1}^N m_i |\bv_i-\bv_0| }.
\end{equation}  
This makes the interaction less ``point-like'', but it provides a more correct
estimate of the scattering rates and their dependence on particle energy.  
We have run one simulation ($\hat{\sigma}_{DM}=1$) with $8$ times as many
particles to estimate the systematic error induced by this non-locality,
and find that it is negligible (see Fig.~\ref{fig:core}a).

Once a particle is selected for the interaction, we assume that the scattering 
is isotropic.  For particles with center of mass velocity $\bv_c$ and velocity
difference $\Delta \bv$, we select a random direction $\be$ and assign the
particles velocities of
\begin{eqnarray}
\bv_0 &=& \bv_c + |\Delta \bv| \be m_j/(m_0+m_j)\\
\bv_j &=& \bv_c - |\Delta \bv| \be m_0/(m_0+m_j). 
\end{eqnarray}
Such elastic scattering will conserve energy and linear momentum, but not
angular momentum because of the finite particle separations.
However, the net violation of angular momentum conservation over all
interactions will be zero because the particle separation vectors have random
orientations.  
The finite interaction range also leads to diffusion effects, but these
introduce negligible error as we can see by comparing to our higher resolution
simulation (Fig.~\ref{fig:core}a).

Finally we note that several simulations have been run approximating
self-interacting dark matter as a fluid and using numerical hydrodynamic
techniques (Moore et al.~\cite{MGJPQ}; Yoshida et al.~\cite{YSWT}).
This is a physically different regime from the low optical depth system
proposed by Spergel \& Steinhardt~(\cite{SpeSte}).
Unlike a strongly interacting fluid, the energy transfer in SIDM is dominated
by stochastic collisions in the core which rapidly transport the energy into
the outskirts of the halo.  The fluid results are thus not directly applicable.
The physical problem SIDM most closely resembles is 2-body relaxation in
globular clusters
(see Spergel \& Steinhardt~\cite{SpeSte}; Hannestad~\cite{Hann};
Burkert~\cite{Bur}).

\begin{figure}
\begin{center}
\leavevmode
\epsfxsize=3.5in \epsfbox{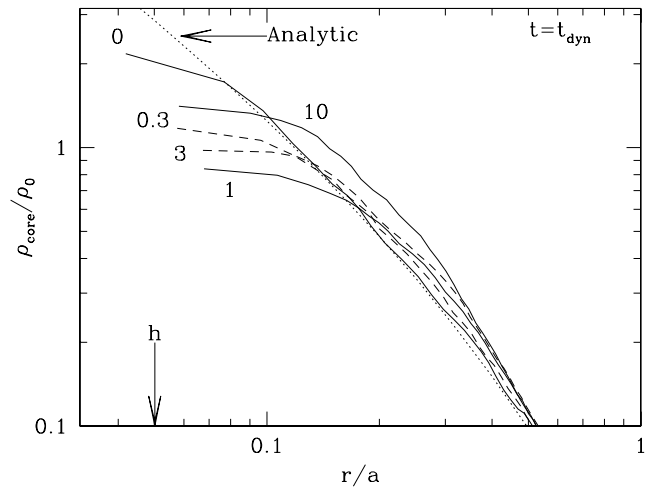}
\end{center}
\caption{(Top) The halo density profiles at $t\simeq \tdyn$ for models with
$\hat{\sigma}_{DM}=0$, $0.3$, $1.0$, $3.0$ and $10.0$. The dotted line shows 
the initial Hernquist profile. Note that after only one dynamical time the
$\hat{\sigma}_{DM}=3$ and $10$ cases have started to core collapse. The 
force softening is indicated by $h$.  } 
\label{fig:profile}
\end{figure}

\begin{figure}
\begin{center}
\leavevmode
\vspace{-0.3in}
\epsfxsize=3.3in \epsfbox{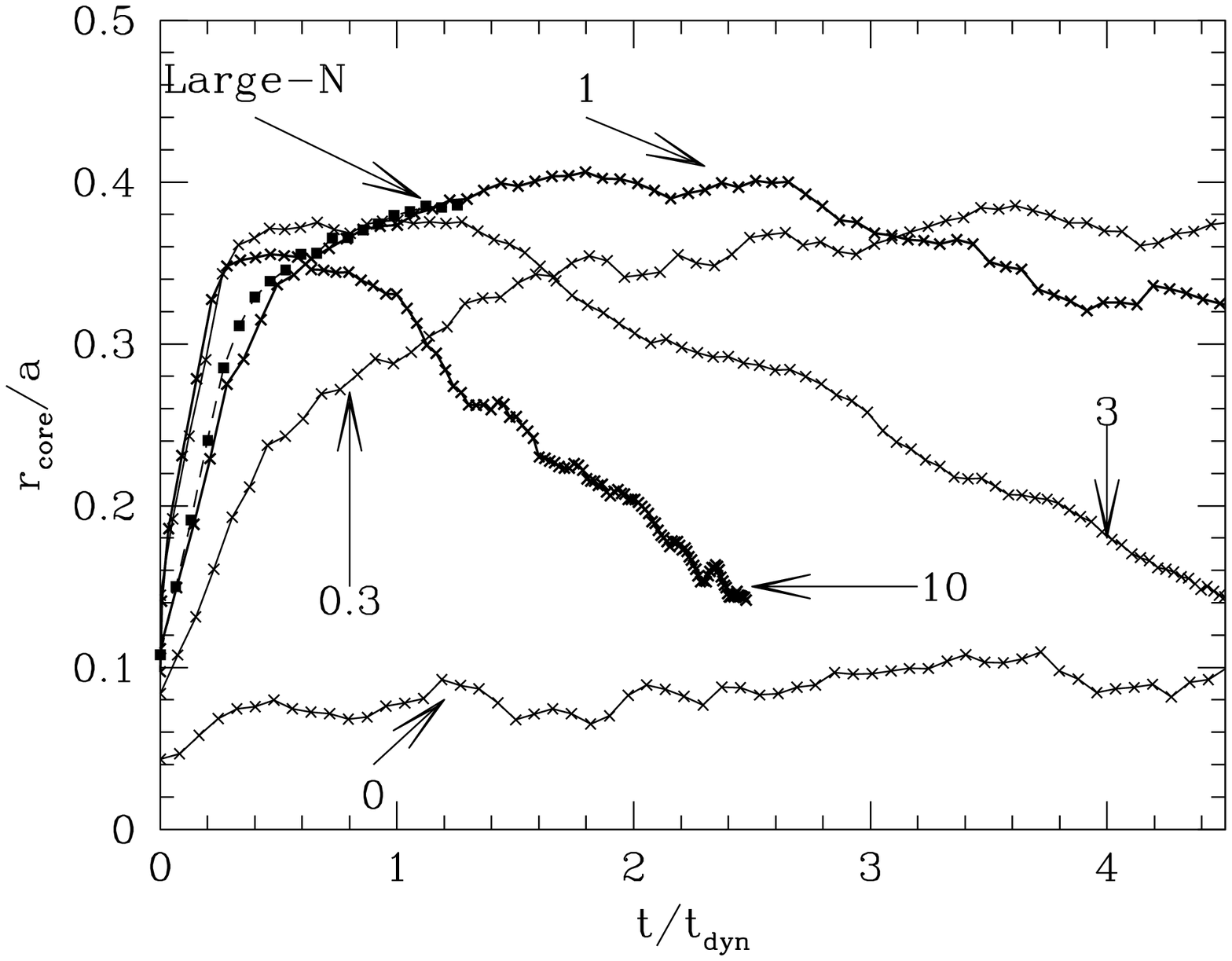}
\end{center}
\begin{center}
\leavevmode
\vspace{-0.3in}
\epsfxsize=3.3in \epsfbox{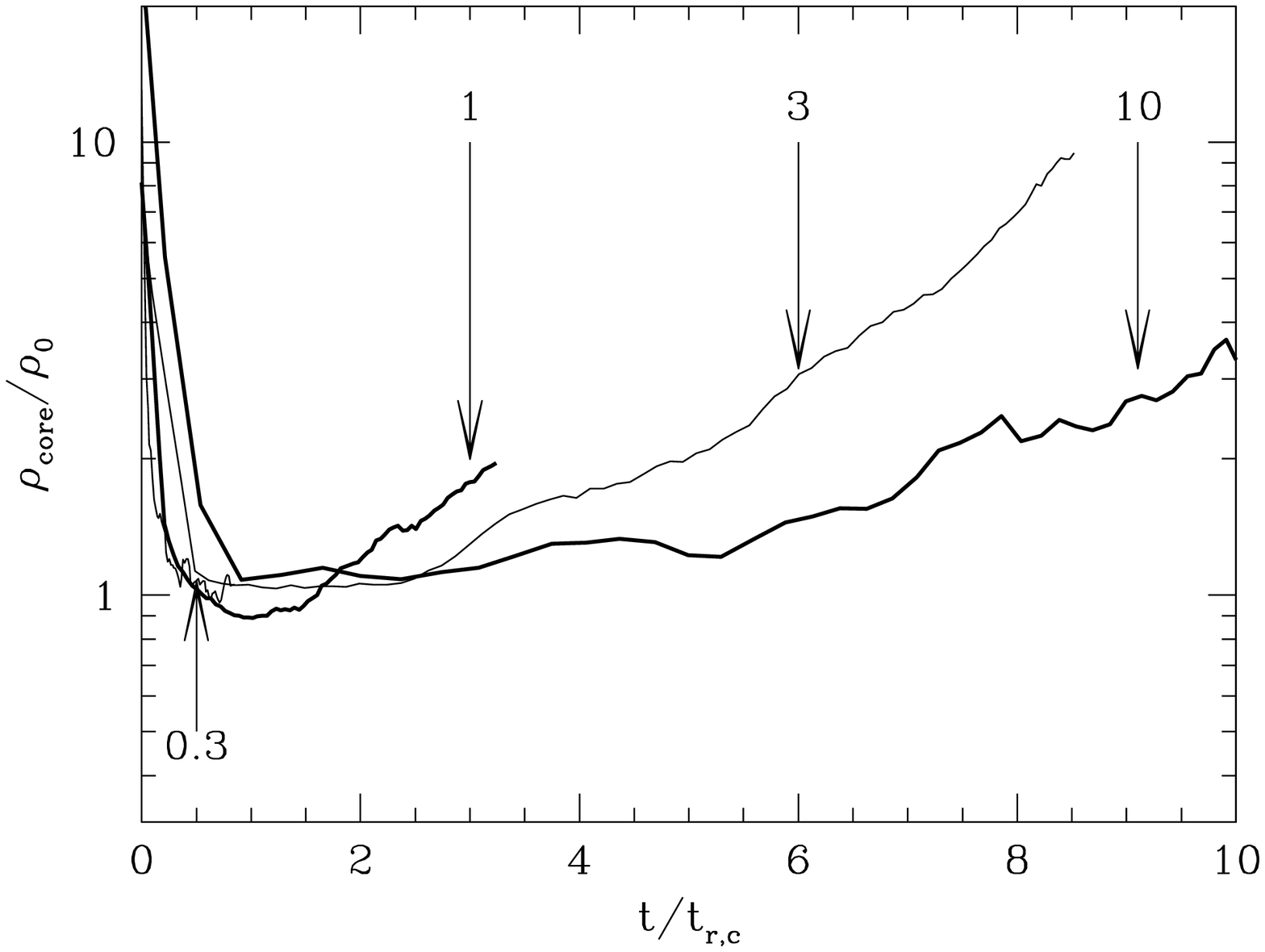}
\end{center}
\begin{center}
\leavevmode
\vspace{-0.3in}
\epsfxsize=3.3in \epsfbox{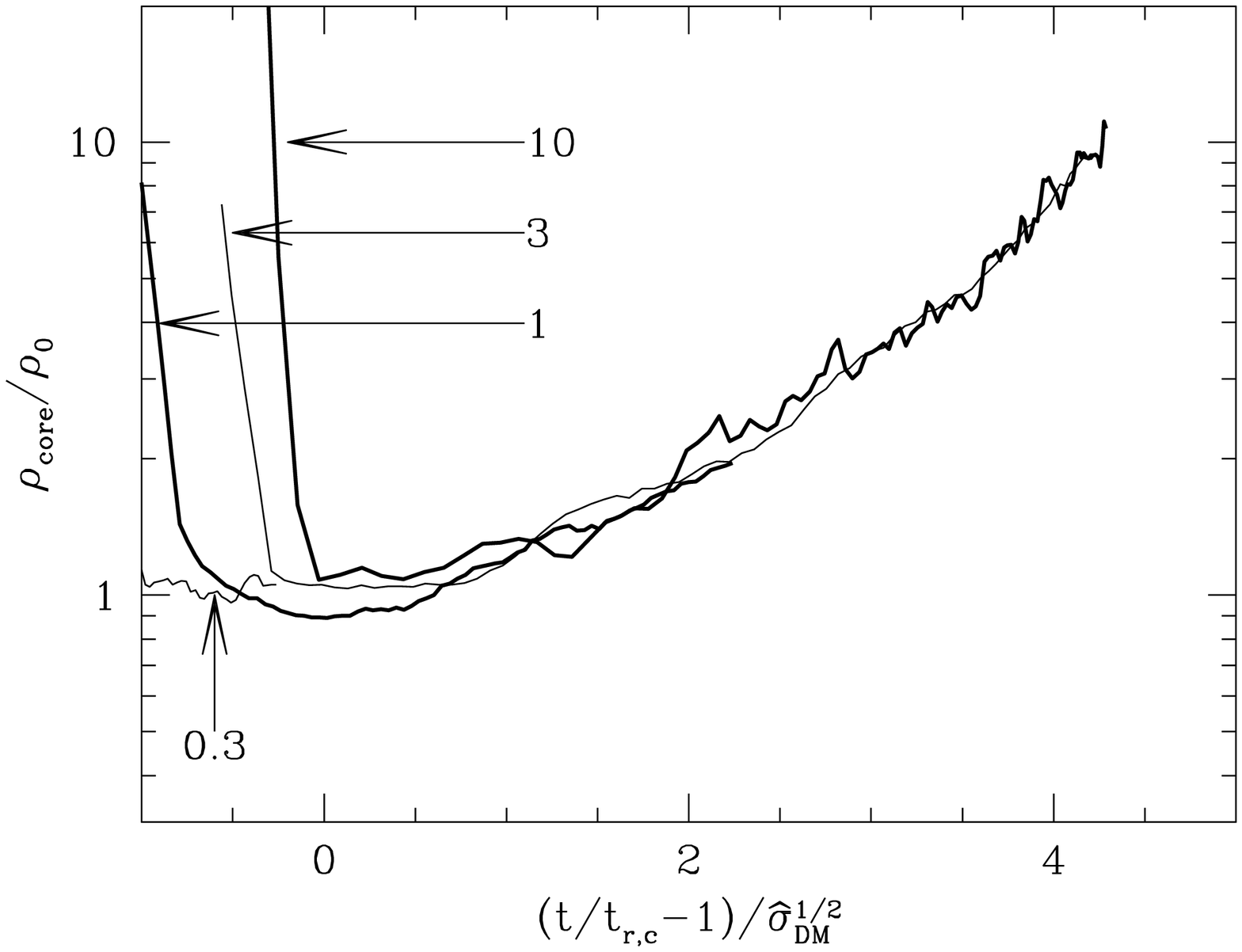}
\end{center}
\caption{(Top) Core radius evolution as a function of dynamical time.
The solid lines show the evolution of the core radius (temporally smoothed)
for dimensionless cross sections of
$\hat{\sigma}_{DM}=0$, $0.3$, $1.0$, $3.0$ and $10.0$.  Note that the standard
($10^5$ particles) and ``Large-N'' ($8\times 10^5$ particles) simulations for
$\hat{\sigma}_{DM}=1$ show identical evolution.
(Middle) Central density evolution as a function of relaxation time
$t_{rc}=1.7\tdyn/\hat{\sigma}_{DM}$.   As expected, the minimum core
density occurs at $t_{rc}\simeq 1$. 
(Bottom) Recollapse.  The recollapse of the core after reaching maximum
expansion near $t/t_{rc}=1$ is self-similar for a time scale of
$t_{rc}\hat{\sigma}_{DM}^{1/2}$. }
\label{fig:core}
\end{figure}

\section{Results} \label{sec:core}

We considered a series of experiments comparing the evolution of N-body systems
with and without SIDM parameterized by $\hat{\sigma}_{DM}$.  The key to 
understanding their evolution is the relaxation time due to the scattering.
It is, however, a different regime from normal 2-body relaxation. The
scattering process is direct rather than diffusive and the structure of the
system evolves on the relaxation time scale rather than on a large multiple of 
the relaxation time scale as in globular clusters.
We first define a relaxation time for SIDM, and then compare the evolution
of three dynamical systems with and without SIDM.  We begin with the
evolution of Hernquist profile galaxies, next we consider the effects of
adiabatic compression of the dark matter by the baryons in modifying this
evolution, and finally we examine the collapse of a cold top-hat perturbation.

\subsection{Relaxation time}

For 2-body relaxation, the relaxation time is defined to be
$\sigma^2/D(v_{\rm para}^2)$ where $\sigma$ is the one-dimensional velocity
dispersion and $D(v_{\rm para}^2)$ is the diffusion coefficient parallel to
the particle velocity (see Binney \& Tremaine~\cite{BinTre}; \S 8).
For SIDM, the scattering angle average of the velocity change is just
${1\over 2}|\Delta \bv|^2$ so the analogy of the 2-body relaxation diffusion
coefficient is 
\begin{eqnarray}
  D(v^2) &=& { 1\over 2} \int \rho\sigma_{DM}|\Delta \bv |^2 
             f(\bv_0)f(\bv_1)d^3\bv_0 d^3\bv_1  \\
         &=& { 16 \over \sqrt{\pi} } \rho \sigma_{DM} \sigma^3 
\end{eqnarray}  
assuming that the particle velocity distributions $f(\bv)$ are Maxwellians
with one-dimensional velocity dispersions of $\sigma$.
The relaxation time is then
\begin{equation}
  t_r \equiv { 3 \sigma^2 \over D(v^2) }
      \simeq { 1 \over 3 \rho \sigma_{DM} \sigma } 
\end{equation}
where the factor of $3$ arises because we considered the total velocity rather
than a single component.
The relaxation time associated with the mean interaction radius $a/2$, which
we will call the core relaxation time, is $t_{rc}=1.7\tdyn/\hat{\sigma}_{DM}$.
At the Hernquist break radius the relaxation time is
 $8\tdyn/\hat{\sigma}_{DM}$ and at the half-mass radius ($2.4a$) it is
$35\tdyn/\hat{\sigma}_{DM}$.
As expected the evolution seems to occur on the relaxation time scale
associated with the mean interaction radius $a/2$.

Unlike globular clusters, the $\rho\propto 1/r$ cusped density distributions evolve rapidly,
and core collapse occurs after only 5 half-mass relaxation times 
(Quinlan~\cite{Qui}).
The 2-body relaxation results will significantly overestimate the actual
core collapse time scale because the SIDM scattering is not diffusive.  A
particle undergoes one scattering with $\Delta v \sim v$ rather than slowly
diffusing in energy due to many small scatterings.  Crudely, we expect the
system to form a constant density core of size $\simeq a/2$ on the time scale
$t_{rc}$ and core collapse faster than the 2-body relaxation prediction of
$5$ half-mass relaxation times.
After core collapse, the system will have a steeper $\rho \propto 1/r^2$ cusp.
This is just the behavior we see in the simulations, as we now describe.

\subsection{Core formation}

We first examined the formation of a core and the alteration of rotation
curves as a function of the dimensionless cross section $\hat{\sigma}_{DM}$.
Fig.~\ref{fig:profile} shows the density distributions 
at $t\simeq\tdyn$ for all five simulations.
The scattering clearly produces a constant density core, while the simulation
without any scattering roughly maintains the input density cusp.
However, even after such a short time interval, the $\hat{\sigma}_{DM}=10$
simulation has begun its core collapse phase and has a higher central density
than any of the other SIDM cases. 
Our results are very similar to those of Burkert~(\cite{Bur}), but the higher
scattering rates found with the better approximation to the scattering rate
integral mean that we obtain the same evolution with approximately 1/3 the
SIDM cross section.

We non-parametrically estimate the core radius by the point where the density
drops to $1/4$ the central density. The central density is found by 
finding the region over which the Kolmogorov-Smirnov test provides a 
reasonable likelihood that the particle distribution is consistent with
a constant density.  This tends to provide us with an upper limit to the
core radius, with an accuracy that is limited primarily by the finite particle
number at very small radius.
Fig.~\ref{fig:core} shows the evolution of the core radius with time and the
central density with core relaxation time $t_{rc}$.
For comparison, the core radius due to the finite particle number and numerical
relaxation effects starts on the force smoothing scale ($0.05a$) and slowly
evolves to $0.1a$, which represents a negligible systematic error for the much
larger core radii created by the SIDM (recall our estimate of the core radius
is in fact an upper limit).
We also find that the evolution of the $\hat{\sigma}_{DM}=1$ system with
$8\times 10^5$ particles is identical to the standard simulation with
$10^5$ particles (see Fig.~\ref{fig:core}a).  

The system evolution is remarkably rapid, with a very narrow window between
the formation of a significant core radius and its recollapse.  The maximum
core size is approximately $0.4a$ and it forms on the core relaxation time
scale $t\simeq t_{rc}=1.7\tdyn/\hat{\sigma}_{DM}$.  Almost immediately after
reaching its maximum size the core begins to collapse again, leaving a brief
window where the dark matter shows a large core radius.  
Although the formation of the core seems to be nearly independent of 
$\hat{\sigma}_{DM}$ when we scale the time by the core relaxation time
scale, the subsequent core collapse phase is not.  The rate of recollapse 
in units of $t_{rc}$ appears to accelerate as the cross section decreases 
(see Fig.~\ref{fig:core}b).  Empirically we find the evolution of the
central density after the reaching the maximum core radius is self-similar
if we scale the time in units of $t_{rc}\hat{\sigma}_{DM}^{1/2} \propto \hat{\sigma}_{DM}^{-1/2}$
(see Fig.~\ref{fig:core}c), which is a different scaling from that for normal 
2-body relaxation where the collapse depends only on the relaxation time ($\propto\hat{\sigma}_{DM}^{-1}$). 
We hypothesize that the difference arises because the scattering is not a
diffusive process.  In each scattering $\Delta v\sim v$ and particles escape
the core in a single scattering, leading to faster evolution.

\subsection{Adiabatic Contraction}

\begin{figure}
\begin{center}
\leavevmode
\epsfxsize=3.5in \epsfbox{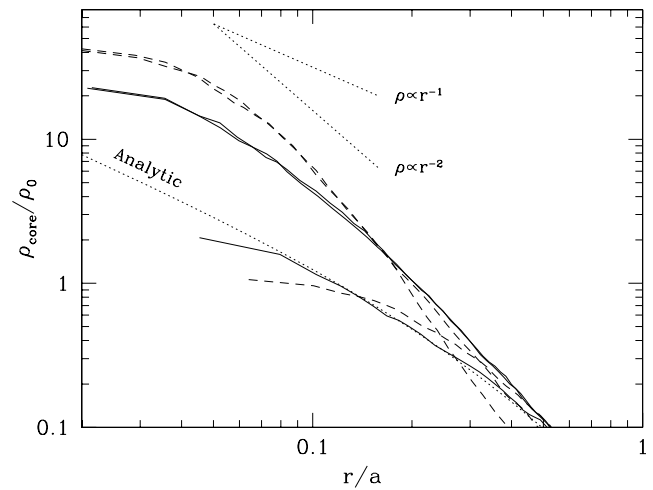}
\end{center}
\caption{The combined effects of SIDM and adiabatic compression.
The SIDM cross section is $\hat{\sigma}=0$ (solid) or $1$ (dashed) and the
adiabatically compressed curves include the baryonic contribution
(Eq.~\protect\ref{eqn:baryons}).  Curves of increasing central density
are just before the compression ($t=\tdyn$), just after the compression
($t=3\tdyn$) and some time after compression ($t=4.0\tdyn$). The
force smoothing scale is $0.01a$.}
\label{fig:adcom}
\end{figure}

Very careful tuning of the SIDM cross section and ages might allow dwarf 
galaxies to have softened cores today without undergoing core collapse.
However, we must examine the effects of the baryons on the dark matter profile
to estimate the appropriate cross section, as
the cooling baryons will adiabatically compress the SIDM dark matter
and significantly modify the evolution of the system.
The adiabatic compression both raises the scattering rates and reduces the
relaxation time scales in three ways.  First, by raising the densities and 
reducing the dynamical time scale, the relaxation time drops. 
Very crudely, if we compress the dark matter by a factor of $f$, the
scattering rate jumps by a factor of $f^{7/2}$, producing a corresponding
reduction in the relaxation time. 
A second, more subtle factor is the difference in how density cusps and
density distributions with cores are adiabatically compressed.  Standard
adiabatic compression models for a $1/r$ density cusp will maintain a 
$1/r$ density cusp after the compression
(see Quinlan, Hernquist \& Sigurdsson~\cite{QHS})
while the compression of a density distribution with a finite core radius
produces a steeper ($1/r^{3/2}$ to $1/r^{9/4}$) cusp depending on the 
distribution function (Young~\cite{Young}).  Thus, if the SIDM produces a
core radius before the adiabatic compression phase, the adiabatic compression
produces a steeper than normal central density cusp.  In 2-body relaxation,
steep cusps evolve more rapidly than shallow cusps, and Quinlan~(\cite{Qui})
found that where a $1/r$ density cusp reaches core collapse in 5 half-mass
relaxation times, a $1/r^2$ density cusp reaches core collapse in 1 half-mass
relaxation time.  Hence the steeper central cusp created by adiabatic 
compression of the SIDM core further accelerates the ultimate evolution of the
system to a central $1/r^2$ density cusp.  Third, the existence of the
baryonic matter provides a component of the gravitational potential whose
evolution is not driven by scattering.

We investigate this with a single, somewhat contrived experiment.  We put
20\% of the mass of the profile in an analytic Hernquist potential with a
corresponding reduction in the masses of the dark matter particles.
We modeled the adiabatic compression by running the system for $\tdyn$, and
then slowly changing the external potential to
\begin{equation}
  \phi(r) = {-GM_b\over\sqrt{r^2+a_d^2}}
  \label{eqn:baryons}
\end{equation}
between $\tdyn$ and $2\tdyn$, finally fixing the external potential for the
further evolution.  We used $M_b=0.2M_T$, $a_d=0.1a$, and reduced the 
force smoothing to $0.01a$.
The final external potential was chosen to have the same circular velocity
profile as a Kuzmin disk (Binney \& Tremaine~\cite{BinTre}), and underestimates
the amount of compression compared to the standard approximation
(Blumenthal et al.~\cite{BFFP}).
It corresponds to a baryonic density
\begin{equation}
  \rho(r) = {3M_b\over 4\pi} {a_d^2\over (r^2+a_d^2)^{5/2}}.
\end{equation}
At the end of the compression phase, the central dark matter density has
risen by a factor of $\sim 10$.  We ran the compression sequence with either
$\hat{\sigma}_{DM}=0$ or $1$. 

Fig.~\ref{fig:adcom} shows the dark matter density and rotation curve profiles
of the adiabatically compressed systems both at the end of the compression 
phase ($3\tdyn$) and the end of the experiment ($4\tdyn$).  At the start
of the compression, the scattering has begun to establish the typical SIDM 
core structure.  The adiabatic compression then produces significantly
different central density profiles.  The no scattering simulation maintains
a roughly $\rho \propto 1/r$ profile shifted upwards in density, while
the $\hat{\sigma}_{DM}=1$ simulation has formed a steeper, nearly 
$\rho \propto 1/r^2$ profile with a {\it higher} central dark matter
density than if there had been no scattering, just as we expected from
the differences in how the two profiles would be adiabatically compressed.  
In the subsequent evolution, the no scattering simulation maintains a stable profile
while the $\hat{\sigma}_{DM}=1$ simulation has a rising central density.  The
system appears to proceed directly into the core-collapse phase of its evolution
despite the fixed baryonic potential.

\subsection{Spherical Collapse}

Our previous experiments assumed that no significant scattering occurred during
the initial collapse of the dark matter halo so that the initial conditions
contained a $\rho \propto 1/r$ density cusp.  We also considered how the
formation of the halo is altered in the SIDM model by considering the very
simple case of the collapse of a top-hat overdensity with vacuum boundary 
conditions both with and without the DM scattering.
These simulations are meant only to illustrate the primary effects of DM
scattering on collapsing perturbations, and not as realistic models of the
formation of galaxies or clusters in a cosmological context. 

We generated a $10^5$ particle uniform density sphere with velocity dispersion
$1\%$ of the virial velocity and allowed it to collapse.
Without particle scattering the collapse formed a profile akin to the Hernquist
profiles of the previous sections, with some of the particles being ejected
from the system.
By fitting a Hernquist profile to the resulting system we identified the
scattering rate appropriate to $\hat{\sigma}_{DM}\simeq1$ and re-ran the simulation
from the same initial conditions including scattering.

The simulations indicate that the final equilibrium profile is similar with
and without scattering, however at fixed dynamical time the scattering run
always has a higher central density.  We postulate that the scattering 
provides a mechanism, similar to dissipation, which allows the system to
attain higher phase space densities than are possible in a purely 
collisionless/dissipationless collapse.  


\section{Conclusions} \label{sec:conclusion}

We combined an N-body code with a Monte Carlo model of scattering to examine
the effects of self-interacting dark matter (SIDM) on the properties of dark
matter halos.  Our models form a one parameter family with ``standard'' CDM
as a limiting case.

We first examined the time evolution of Hernquist models with $1/r$ central 
density cusps in the presence of SIDM. Our simulations are similar to those
of Burkert~(\cite{Bur}), but use a better approximation to the scattering
integrals.  The models form a constant density core with a
radius $\sim 0.4 a$ on the core radius relaxation time scale
$t_{rc}=1.7\tdyn/\hat{\sigma}_{DM}$, where the SIDM cross section per unit mass
is $\hat{\sigma}_{DM}=\sigma_{DM} M_T/a^2$.
The initial growth of the core radius is very rapid because the relaxation
time is much shorter in the inner regions (see Quinlan~\cite{Qui}).
Shortly after reaching this maximum core size the core begins to shrink.
We find that the recollapse proceeds on a faster time scale than expected
from 2-body relaxation. The central density evolution evolves on
time scales of $t_{rc}\hat{\sigma}_{DM}^{1/2}\propto \hat{\sigma}_{DM}^{-1/2}$ 
rather than $t_{rc}\propto\hat{\sigma}_{DM}^{-1}$, probably because SIDM is 
not a diffusive process when the mean free paths are large.  Thus,
the dark matter briefly has a large, finite core radius which could be fine 
tuned to address the problem of dwarf galaxy rotation curves at the expense 
of predicting core collapsed dark matter halos with $\rho \propto 1/r^2$ cusps 
in most other galaxies. In particular, it is probably impossible to use SIDM 
to eliminate the dark matter cusps in the dwarf galaxies while preserving them 
in the low surface brightness galaxies.

Even with SIDM, however, we must also consider the role of the baryons in
modifying the structure of the dark matter halo.
In particular, as the baryons cool and form a disk they adiabatically compress
the dark matter (Blumenthal et al.~\cite{BFFP}; Dubinski~\cite{Dub}).
The adiabatic compression raises the central dark matter density, increases
the SIDM scattering rate, and reduces the time scale for core collapse.
Moreover, adiabatic compression of a density profile with a finite core
radius produces a steep central density cusp (Young~\cite{Young}) which
relaxes to form a $\rho \propto 1/r^2$ density cusp even faster than
the shallower $\rho \propto 1/r$ cusp we considered initially
(Quinlan~\cite{Qui}).
As a result, the development of a core radius due to the SIDM scattering is
first reversed by the compression, and then core collapse begins on a far shorter time
scale.  Such a further acceleration of the time scale for producing a final
density distribution with a steep $\rho \propto 1/r^2$ density cusp will
make it even harder for SIDM to maintain a core radius in the
dark matter profile for periods comparable to the presumed lifetimes of
dwarf galaxies.

All of the above simulations supposed that the halo collapsed and virialized
before scattering became important, which is appropriate to ``low'' scattering
rates.  Finally, we considered the effects of SIDM on the collapse of cold
top-hat perturbations with $\hat{\sigma}$ large enough that scattering and
virialization could occur at the same time.  We found that the scattering
allowed higher phase space densities at fixed dynamical time than could be
found in the collisionless systems.

It thus appears as though SIDM exacerbates rather than solves the ``central
density cusp problem''.  Turning this around we can say that astrophysics 
almost certainly requires SIDM cross sections small enough to avoid significant
scattering over the age of the universe for all galaxies, or mean free paths
$\ga 1$~Mpc.

\bigskip
\acknowledgments
We would like to acknowledge useful conversations with A.~Burkert and
L.~Hernquist.  M.W. thanks J.~Bagla and V.~Springel for numerous helpful conversations
on N-body codes.  M.W. was supported by NSF-9802362, and C.S.K. was
supported by the Smithsonian Institution.

\appendix

\section{Tree N-body Code} \label{sec:tree}

We have used a new implementation of a {\sl Tree\/} code to evolve our dark
matter halo.  Since this code is new we briefly describe it here, and also
explain how we generated the initial conditions.
 
The basic algorithm is described in Barnes \& Hut~(\cite{BarHut}).
Space is partitioned into an oct-tree structure by bisection in the three
spatial dimensions.  Starting at the root node, which is the entire volume
of the simulation, the tree descends until each node contains at most one
particle.  The force on any given particle from all of the other particles
is computed by a tree walk.  The speed-up over direct summation is obtained
by imposing an opening criterion during the tree walk.  For cells sufficiently
far from the particle in question, the entire mass distribution within that
cell is approximated by a point at the center of mass of the cell and with
mass equal to the sum of the particle masses within the daughters of that cell.
 
Though the code does not require it, we have used equal mass particles
throughout.  The size of the computational box grows so as to always encompass
all of the particles.
Rather than a Plummer potential we use a spline softened force
(Monaghan \& Lattanzio~\cite{MonLat}; Hernquist \& Katz~\cite{HerKat};
Springel, Yoshida \& White~\cite{SprYosWhi}).  For most of the runs we use
$h=0.05a$ (the force was therefore exactly $1/r^2$ beyond $h$), but for the
adiabatic compression runs we lower this to $0.01a$.
Very roughly this corresponds to a Plummer law smoothing $\epsilon\simeq h/3$,
although a Plummer law gives 1\% force accuracy only beyond $10\epsilon$.

With $10^5$ particles it is finite particle numbers which limit our
resolution, {\it not\/} force accuracy.
For profiles with almost any central cusp, the relaxation time is less
than $10\tdyn$ for radii $r<0.05a$ (Quinlan~\cite{Qui}).
Our numerical results with differing particle numbers and force accuracies
support the scalings in Fig.~2 of Quinlan~(\cite{Qui}).
Only for the adiabatic compression runs, where more particles are concentrated
at small radius and there is a stable external potential, do we gain by
lowering $h$.
 
The time integration is done with a second order leap-frog method.
The time step is chosen to be the smaller of $0.5\sqrt{h/a_{\rm max}}$,
where $a_{\rm max}$ is the maximum acceleration on the particle, or
$2h/v$ where $v$ is the particle velocity.  We have shown that this time
step accurately integrates binary orbits, and in combination with the
opening criterion described below conserves total energy to much better
than a percent over 5 dynamical times, except during the core collapse
phase where the error rises to a few percent.
 
Finally for the tree walk we do not simply use the geometrical criterion
advocated by Barnes \& Hut.
Rather we have augmented it with a modification of the method described in
Springel et al.~(\cite{SprYosWhi}).
Starting with the known accelerations from the last time step we descend the
tree.  We open an internal node if the partial acceleration from the
quadrupole moment of that node exceeds $\alpha$ times the old acceleration,
where $\alpha$ is a tolerance parameter.
Specifically if $\ell$ is the size of a cell containing mass $M$ whose center
of mass is a distance $r$ from the particle in question, we open the node if
\begin{equation}
  M \ell^2 > \alpha\left|\vec{a}_{\rm old}\right| r^4
  \qquad .
\end{equation}
This ensures that the relative force accuracy is ${\cal O}(\alpha)$ while the
resulting tree walk is significantly faster (Springel et al.~\cite{SprYosWhi}).
The tree walk is also used to define the neighbor list for the density and
scattering calculations.  To ensure that all nearby cells are opened so the
neighbor list is ``complete'' we augment the opening criterion to additionally
require that $\ell<\theta r$ for cells closer than $a$.
This provides a good estimate of the density except in the lowest density
regions where the scattering probability is anyway small.
Throughout we have used $\alpha=2\%$ and $\theta=0.4$, which for a Hernquist
profile results in a 90th percentile relative force error of $5\times 10^{-3}$,
i.e.~90\% of the particles have force errors less than this.

We set up our initial conditions as a Hernquist~(\cite{Her}) profile
\begin{equation}
  \rho_H(r) = {M_{\rm halo}\over 2\pi} {a\over r} {1\over(r+a)^3}
\end{equation}
where $a$ is the scale-length.
First we choose the position of a particle with a radial probability
distribution proportional to $r^2\rho(r)$ and an isotropic $\hat{r}$.
We impose an upper radial distance of $r_{\rm max}=50a$ at which point
the density is $\sim 10^{-5}$ of the density at $r=a$.
Given the position we then calculate the magnitude of the velocity from
the known distribution function $f(E)$ for this model using an
acceptance-rejection method and again distribute $\hat{v}$ isotropically.
This procedure is iterated $N_p/2=5\times 10^4$ times.
Each particle and a reflected counterpart at $-\vec{r}$ with velocity
$-\vec{v}$ is added to the list to obtain a realization of the halo with
$N_p$ particles.  The use of `reflected' initial conditions ensures that
the center of mass position and velocity of the halo are both initially zero.


\begin{thebibliography}{99}
\bibitem[1986]{BarHut}
Barnes J., Hut P., 1986, Nature, 324, 446
\bibitem[2000]{Bento}
Bento, M.C., Bertolami, O., Rosenfeld, R., \& Teodoro, L., 2000, 
  preprint [astro-ph/0003350]
\bibitem[2000]{BinGer}
Binney J., Gerhard, O., \& Silk, J., 2000, preprint [astro-ph/0003199]
\bibitem[1987]{BinTre}
Binney J., Tremaine S., 1987, ``Galactic Dynamics'', Princeton UP
\bibitem[1986]{BFFP}
Blumenthal G.R., Faber S.M., Flores R., Primack J.R., 1986,
  \apj, 301, 27
\bibitem[2000]{BKW}
Bullock J., Kravstov A.V., Weinberg D.H., 2000, preprint [astro-ph/0002214]
\bibitem[2000]{Bur}
Burkert A., 2000, preprint [astro-ph/0002409]
\bibitem[1994]{CMH}
Carlson E.D., Machacek M.E., Hall L.J., 1992, \apj, 398, 43
\bibitem[1994]{Dub}
Dubinski J., 1994, \apj, 431, 617
\bibitem[1997]{Faber}
Faber, S.M., Tremaine, S., Ajhar, E.A., Byun, Y.-I., Dressler, A.,
  Gebhardt, K., Grillmair, C., Kormendy, J., Lauer, T.R., \& Richstone, D.,
  1997, \aj, 114, 1771
\bibitem[2000]{FOACH}
Firmani C., et al., 2000, preprint [astro-ph/0002376]
\bibitem[1994]{FloPri}
Flores R., Primack J.R., 1994, \apj, 427, L1
\bibitem[1999]{GSL}
Gelato, S. \& Sommer-Larsen, J., 1999, MNRAS, 303, 321
\bibitem[2000]{Hann}
Hannestad S., 2000, preprint [astro-ph/9912558]
\bibitem[2000]{HanSch}
Hannestad S., Scherrer R.J., 2000, preprint [astro-ph/0003046]
\bibitem[1990]{Her}
Hernquist L., 1990, \apj, 356, 359
\bibitem[1989]{HerKat}
Hernquist L., Katz N., 1989, \apjs, 70, 419
\bibitem[1992]{HerW}
Hernquist, L. \& Weinberg, M.D., 1992, ApJ, 400, 80
\bibitem[2000]{HogDal}
Hogan C.J., Dalcanton J.J., 2000, preprint [astro-ph/0002330]
\bibitem[2000]{Hozumi}
Hozumi, S., Burkert, A. \& Fujiwara, T., 2000, MNRAS, 311, 377
\bibitem[1999]{Huss}
Huss A, Jain B., Steinmetz M., 1999, \apj, 517, 64
\bibitem[1999]{Kly}
Klypin A., Kravstov A.V., Valenzuela O., Prada F., 1999, \apj, 522 82
\bibitem[1998]{Kravstov}
Kravstov A.V., et al., 1998, \apj, 502, 48
\bibitem[1999]{Kull}
Kull A., 1999, \apj, 516, L5
\bibitem[1995]{deLaix}
de Laix A.A., Scherrer R.J., Schaeffer R.K., 1995, \apj, 452, 495
\bibitem[1999]{Maclow}
MacLow, M.-M., \& Ferrara, A., 1999, ApJ, 513, 142
\bibitem[2000]{MoMao}
Mo H.J., Mao S., 2000, preprint [astro-ph/0002451]
\bibitem[1985]{MonLat}
Monaghan J.J., Lattanzio J.C., 1985, Astr Ap, 149 135
\bibitem[1994]{Moo94}
Moore B., 1994, Nature, 370, 629 [astro-ph/9402009]
\bibitem[1998]{Moo98}
Moore B., Governato F., Quinn T., Stadel J. Lake G., 1998, \apj, 499, L5
\bibitem[2000a]{Moo99}
Moore B., Quinn T., Governato F., Stadel J., Lake G., 2000a, preprint
  [astro-ph/9903164]
\bibitem[1999]{MGGLQST}
Moore B., et al., 1999, \apj, 524, L19
\bibitem[2000b]{MGJPQ}
Moore B., et al., 2000b, preprint [astro-ph/0002308]
\bibitem[1996]{NEF}
Navarro J., Eke, V.R., \& Frenk C.S., 1996, \mnras, 283, L72
\bibitem[1997]{NFW}
Navarro J., Frenk C., White S.D.M., 1997, \apj, 490, 493
\bibitem[2000]{Ost}
Ostriker J.P., 2000, preprint [astro-ph/9912548]
\bibitem[1995]{QHS}
Quinlan G.D., Hernquist, L., \& Sigurdsson, S., 1995, ApJ, 440, 554
\bibitem[1996]{Qui}
Quinlan G.D., 1996, NewA, 1, 255
\bibitem[2000]{SpeSte}
Spergel D.N., Steinhardt P.J., 2000, preprint [astro-ph/9909386]
\bibitem[2000]{SprYosWhi}
Springel V., Yoshida N., White S.D.M., 2000, preprint [astro-ph/0003162]
\bibitem[1998]{SyeWhi}
Syer D., White S.D.M., 1998, \mnras, 293, 337
\bibitem[2000]{vdBRobDalBlo}
van den Bosch F.C., Robertson B.E., Dalcanton J.J., de Blok W.J.G., 2000,
  \aj, in press [astro-ph/9911372]
\bibitem[2000]{YSWT}
Yoshida N., Springel V., White S.D.M., Tormen G., 2000, preprint
  [astro-ph/0002362]
\bibitem[1980]{Young}
Young, P., 1980, \apj, 242, 1232
\end{thebibliography}
\end{document}